\documentclass[aps,pra,twocolumn,superscriptaddress,amsmath,amssymb,tightenlines,epsfig,floatfix]{revtex4}
\usepackage{graphicx}
\usepackage{dcolumn}	
\usepackage{bm}			
\usepackage{amsfonts}
\usepackage{xspace}

\begin{document}

\newcommand{\psihat}{\ensuremath{\hat{\psi}}\xspace}
\newcommand{\psihatd}{\ensuremath{\hat{\psi}^{\dagger}}\xspace}
\newcommand{\ahat}{\ensuremath{\hat{a}}\xspace}
\newcommand{\ahatd}{\ensuremath{\hat{a}^{\dagger}}\xspace}
\newcommand{\bhat}{\ensuremath{\hat{b}}\xspace}
\newcommand{\bhatd}{\ensuremath{\hat{b}^{\dagger}}\xspace}
\newcommand{\boldr}{\ensuremath{\mathbf{r}}\xspace}
\newcommand{\dr}{\ensuremath{\,d^3\mathbf{r}}\xspace}
\newcommand{\etal}{\emph{et al.\/}\xspace}
\newcommand{\ie}{i.e.\:}
\newcommand{\eq}[1]{Eq.\,(\ref{#1})\xspace}
\newcommand{\fig}[1]{Figure\,(\ref{#1})\xspace}
\newcommand{\abs}[1]{\left| #1 \right|} 

\title{Surpassing the Standard Quantum Limit in an Atom Interferometer with Four-mode Entanglement Produced from Four-Wave Mixing.}
\author{S. A. Haine}
\affiliation{Australian Research Council Centre of Excellence for Quantum-Atom Optics}
\affiliation{University of Queensland, Brisbane, 4072, Australia}
\email{haine@physics.uq.edu.au}
\author{A. J. Ferris}
\affiliation{University of Sherbrooke, Canada}
\preprint{Version: submitted \today}

\begin{abstract}
We theoretically investigate a scheme for atom interferometry that surpasses the standard quantum limit. A four-wave mixing scheme similar to the recent experiment performed by Pertot \etal \cite{pertot} is used to generate sub-shot noise correlations between two modes. These two modes are then interfered with the remaining two modes in such a way as to surpass the standard quantum limit, whilst utilising all of the available atoms. Our scheme can be viewed as using two correlated interferometers. That is, the signal from each interferometer when looked at individually is classical, but there are correlations between the two interferometers that allow for the standard quantum limit to be surpassed. 
\end{abstract}

\maketitle

\section{Introduction} 
Atom interferometry \cite{pritchard_review} is a useful tool for sensitive measurements of gravitational fields and accelerations \cite{chu99, chu2001}, gravitational gradients \cite{kasevich02}, rotations \cite{kasevich97}, magnetic fields \cite{stamperkurn07}, the gravitational constant ($G$) \cite{kasevichG}, and the fine structure constant ($\alpha$) \cite{biraben_alpha}. Although most state of the art atom interferometers currently utilise laser cooled atoms, there may be some benefit to using Bose-condensed atoms. Bose-Condensed atoms provide improved visibility in configurations which require complex manipulation of the motional state, such as high momentum transfer beamsplitters \cite{close2011}. The ultimate limit of the sensitivity of any interferometric device which uses uncorrelated particles in each arm is the standard quantum limit $\Delta \phi = \frac{1}{\sqrt{N_t}}$, where $N_t$ is the total number of detected particles \cite{dowling}. 

Recently, there has been much interest in surpassing this limit in atom interferometers via the use of quantum entanglement, with two recent proof of principles experiments demonstrating sub-shotnoise phase sensing \cite{oberthaler2010, treutlein2010}. These experiments exploit the $s$-wave scattering to provide one-axis twisting \cite{ueda, hainekerr} to generate spin squeezing in the two atomic modes. However, both of these experiments use only a small number of atoms ($200$-$1200$ atoms), so the absolute sensitivity of the device is low. It is unlikely that the number of atoms used in this scheme can be increased significantly, as the phase correlations produced become very sensitive to classical uncertainty in the total number of particles as the total number of particles is increased.

Atomic four-wave mixing \cite{phillipsFWM} has long been considered a possible method of creating entanglement between spatially separated atomic modes \cite{ketterleFWM1, ketterleFWM2, ferrisFWM, trippenbach_collision, drummond_collision, ogren2008, dall2009, dennis2010}, and the generation and detection of quantum correlations using this process has recently been demonstrated \cite{aspectFWM, aspectFWM2}. Four-wave mixing has an advantage over one-axis twisting schemes as the correlations produced by the interaction are number correlations, and are not as sensitive to the total number of particles. However, it is so far unclear how best to utilise these correlations for the purpose of atom interferometry. We consider a setup similar to the recent experiment of Pertot \etal \cite{pertot}. Here, collisions between two electronic states allow for collinear four-wave mixing, which can be well represented by a four-mode model. A simple theory predicts that the two modes with low occupation number display sub-shot noise quantum correlations. However, in order to increase the sensitivity of our device, we devise an interferometry scheme which utilises \emph{all} of the available atoms, such that it is possible to surpass the standard quantum limit while retaining a larger number of atoms.  

The structure of this paper is as follows. In section \ref{sec2}, we will describe our proposed scheme. In section \ref{sec3}, we will introduce a simple four-mode model to calculate the level of quantum correlations generated from the four-wave mixing process. In section \ref{sec4} we will introduce a more realistic multi-mode model that allows for effects such as spatial inhomogeneity and multiple order scattering. In section \ref{sec5} we will demonstrate how this system can be used to perform an interferometric phase measurement which surpasses the standard quantum limit, and in section \ref{sec6} we will compare our scheme to one-axis twisting for a large number of particles.

\section{Four-wave mixing. }\label{sec2}
\begin{figure}
\includegraphics[width=0.6\columnwidth]{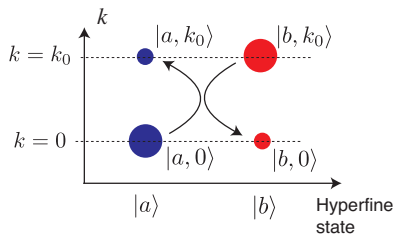}
\caption{\label{fig:fwm_scheme} (Color online) Four-wave mixing. $s$-wave collisions between $|a,0\rangle$ and $|b,k_0\rangle$ scatter atoms into $|a,k_0\rangle$ and $|b,0\rangle$.  }
\end{figure}

Our scheme involves two stages: The \emph{state preparation} stage (Fig. (\ref{fig:fwm_scheme})), and the \emph{interferometry} stage. We begin the state preparation stage with a BEC containing approximately $2\times 10^5$ $^{87}$Rb atoms, tightly confined in the $z$ and $y$ direction, and weakly confined in the $x$ direction. The $x$ confinement is switched off before the state preparation begins. All atoms are in the $|a\rangle \equiv |F=1, m_F =1\rangle$ hyperfine state. To induce the four-wave mixing, we couple approximately $50\%$ of the atoms to the $|b\rangle \equiv |F=2, m_F=0\rangle$ hyperfine state via a two-photon Raman transition, which also transfers linear momentum $\hbar k_0$ to the atoms. In addition, we use resonant microwave coupling to transfer a small amount of population (approximately $1\%$) between $|a\rangle$ and $|b\rangle$ with no momentum transfer. At this stage, modes $|a,0\rangle$ and $|b,k_0\rangle$ each contain approximately $10^5$ atoms, and modes $|b,0\rangle$ and $|a, k_0\rangle$ each contain approximately $10^3$ atoms, where the notation $|a (b),k\rangle \equiv |a (b)\rangle|k\rangle$, and the state $|k\rangle$ indicates a wavepacket traveling with mean momentum $\hbar k$. $s$-wave collisions between $|a,0\rangle$ and $|b,k_0\rangle$ atoms will transfer particles from $|a,0\rangle$ to $|a,k_0\rangle$, and $|b,k_0\rangle$ to $|b,0\rangle$. The creation of one $|a,k_0\rangle$ atom indicates that a $|b,0\rangle$ atom has also been created, as well as the removal of one atom each from modes $|a,0\rangle$ and $|b,k_0\rangle$. This is the source of the quantum correlations. The $1\%$ seed is optional, as the tight transverse confinement strongly suppresses scattering into other modes. However, we found that the addition of a small seed improved the multimode dynamics of the system, as will be discussed in section \ref{sec4}. After an amount of time $t_{fwm}$, the desired level of four-wave mixing has been achieved, and the two momentum modes completely spatially separate, such that at time $t_1$, $|a,0\rangle$ and $|b,0\rangle$ are spatially overlapping, centred at locations $x_L$, and $|a,k_0\rangle$ and $|b,k_0\rangle$ are spatially overlapping, centred at locations $x_R$. $t_{fwm}$ can be adjusted by adjusting the longitudinal and transverse trapping frequencies, which will adjust the ratio of the characteristic four-wave mixing time and the time taken for the two momentum wave packets to spatially separate. Alternatively, a Feshbach resonance could be used to switch off the $s$-wave interactions after an amount of time $t_{fwm}$. At this stage, the system can be thought of as two spin-$\frac{1}{2}$ systems. One, formed by $|a,0\rangle$ and $|b,0\rangle$ centred at $x_L$, and the other formed by $|a,k_0\rangle$ and $|b,k_0\rangle$ centred at $x_R$

\section{Four-mode model}\label{sec3}
In this section we will introduce a simple four-mode model to calculate the degree of quantum correlations generated from the four-wave mixing process. Beginning with a full multimode model describing the system, the Hamiltonian is
\begin{eqnarray}
\mathcal{H} &=& \sum_j \int \psihatd_j(\boldr) H_0 \psihat_j(\boldr) \, \dr  \label{fullham} \\ 
&+& \sum_{i}\sum_{j} \frac{U_{ij}}{2}\int \psihatd_i(\boldr)\psihatd_j(\boldr)\psihat_i(\boldr)\psihat_j(\boldr) \, \dr \, , \nonumber
\end{eqnarray}
where $j = a, b$, and $H_0 = \frac{-\hbar^2}{2m}\nabla^2 + V(y,z)$, and $U_{ij} \equiv \frac{4\pi\hbar^2 a_{ij}}{m}$, where $a_{ij}$ is the inter- and intra- species scattering length, and $m$ is the mass of the atom. We can gain insight into the behaviour of the system by introducing a simple four-mode model by making the approximation
\begin{eqnarray}
\psihat_{a}(\boldr) &\approx& \hat{a}_0\Psi(x,y,z) + \hat{a}_{k_0}\Psi(x,y,z)e^{ik_0 z} \nonumber \\
\psihat_{b}(\boldr) &\approx& \hat{b}_0\Psi(x,y,z) + \hat{b}_{k_0}\Psi(x,y,z)e^{ik_0 z} \, , \label{smode_approx}
\end{eqnarray}
where $\Psi(x,y,z)$ is the (normalized) ground state of the confining potential, before the longitudinal potential was switched off. Substituting this into \eq{fullham} gives
\begin{eqnarray}
\mathcal{H} &=& \sum_{k = 0, k_0}\hbar \omega_k\left(\ahatd_k\ahat_k + \bhatd_k\bhat_k\right) \nonumber  \\
&+&\frac{\hbar\chi}{2}\sum_{k, k^{\prime} = 0, k_0}\left(\ahatd_k\ahatd_{k^\prime}\ahat_k\ahat_{k^\prime} + \bhatd_k\bhatd_{k^\prime}\bhat_k\bhat_{k^\prime} + 2\ahatd_k\ahat_k\bhatd_{k^\prime}\bhat_{k^\prime} \right)  \nonumber \\
&+& \hbar\chi\left(\ahatd_{k_0}\bhatd_{0}\ahat_{0}\bhat_{k_0} + \ahatd_{0}\bhatd_{k_0}\ahat_{k_0}\bhat_{0}\right) \label{4mode_ham}
\end{eqnarray} 
where $\hbar\omega_k = \hbar\omega_0+ \frac{\hbar k^2}{2m}$, and $\hbar\omega_0$ is the ground state eigenvalue of $H_0$. The characteristic four-wave mixing rate is $\chi \equiv U_0\int |\Psi(x,y,z)|^4 \, \dr$, and we have assumed that $U_{aa} =U_{bb}=U_{ab}$, which is a good approximation for $^{87}Rb$. We have assumed that $\int |\Psi(x,y,z)|^2 e^{i k_0 x} \, \dr \approx 0$, which is a good approximation when the momentum spread of the condensate is small $\hbar k_0$, or $k_0 \gg \frac{1}{r_x}$, where $r_x$ is the characteristic size of the condensate in the $x$ direction. 

We can gain some insight into the behaviour of this system with a simple analytic model. The Heisenberg equations of motion for the two lowly occupied `seed' modes are
\begin{eqnarray}
i\dot{\ahat}_{k_0} &=&  \left(\omega_{k_0} + \chi\hat{N}_t \right)\ahat_{k_0} + \chi \ahat_0\bhat_{k_0}\bhatd_0 \label{HEOM1} \\
i\dot{\bhat}_0 &=&  \left(\omega_0 + \chi\hat{N}_t\right)\bhat_0 + \chi \ahat_0\bhat_{k_0}\ahatd_{k_0} \label{HEOM2}
\end{eqnarray}
where $\hat{N_t} \equiv  \ahatd_0\ahat_0+ \bhatd_0\bhat_0+ \ahatd_{k_0}\ahat_{k_0}+ \bhatd_{k_0}\bhat_{k_0}$ is the total number of atoms, which we will treat as a constant $\hat{N_t}\rightarrow N_t$, as it is a conserved quantity. We define the number of particles in each mode as $\hat{N}_{aL} = \ahatd_0\ahat_0$, $\hat{N}_{bL} = \bhatd_0\bhat_0$, $\hat{N}_{aR} = \ahatd_{k_0}\ahat_{k_0}$, and $\hat{N}_{bR} = \bhatd_{k_0}\bhat_{k_0}$. As $\langle \hat{N}_{aL}\rangle, \langle \hat{N}_{bR}\rangle \gg \langle \hat{N}_{aR}\rangle, \langle \hat{N}_{bL}\rangle$ at $t=0$,  we can make the approximation $\ahat_0\rightarrow \alpha_0$, $\bhat_{k_0}\rightarrow \beta_{k_0}e^{-i\omega_{k_0}t}$, where $\alpha_0 = \sqrt{\langle\hat{N}_{aL}\rangle}$, and $\beta_{k_0} = \sqrt{\langle \hat{N}_{bR} \rangle}$. By ignoring the depletion from these modes, which is acceptable for times $\chi N_t t \ll 1$, we find the solution to \eq{HEOM1} and \eq{HEOM2} is
\begin{eqnarray}
\tilde{a}_{k_0}(t) &=& \ahat_{k_0}(0)\cosh r -i \bhatd_0(0)\sinh r \nonumber \\ 
\\
\tilde{b}_{0}(t) &=& \bhat_{0}(0)\cosh r -i \ahatd_{k_0}(0)\sinh r\, , \nonumber \\
\end{eqnarray}
with $r = \chi\alpha_0\beta_{k_0} t$ and we have made the transformation $\tilde{a}_{k_0} = \ahat_{k_0}e^{i\left(\omega_{k_0} +N_t\chi\right)t}$, $\tilde{b}_{0} = \bhat_{0}e^{i\left(\omega_0 +N_t\chi \right)t}$

The quantity we are interested in is the relative number difference variance, which we define as 
\begin{eqnarray}
v_{i,j} &\equiv& \frac{V(N_-)}{\langle \hat{N}_+\rangle} \\
&\equiv& \frac{\langle \left(\hat{N}_i - \hat{N}_j\right)^2\rangle - \langle \left(\hat{N}_i - \hat{N}_j\right)\rangle^2}{\langle \hat{N}_i + \hat{N}_j\rangle} \nonumber
\end{eqnarray}
$v_{i,j} = 1$ for two independent Glauber coherent states \cite{Walls}, which is a good approximation of the quantum statistics of a coherently split BEC. Taking the initial quantum state of modes $|a, k_0\rangle$ and $|b, 0\rangle$ to be independent Glauber coherent states with mean populations $\langle \hat{N}_{aR}(0)\rangle = \langle \hat{N}_{bL}(0)\rangle \equiv N_0$, we find that
\begin{equation}
\langle \hat{N}_{aR}(t) \rangle =\langle \hat{N}_{bL}(t) \rangle= \left(N_0 +\frac{1}{2}\right)\cosh 2r -\frac{1}{2} \, ,
\end{equation}
and
\begin{equation}
v_{aR, bL} = \frac{2N_0}{\left(2N_0 +1\right)\cosh 2r -1} \, .
\end{equation}

\begin{figure}
\includegraphics[width=\columnwidth]{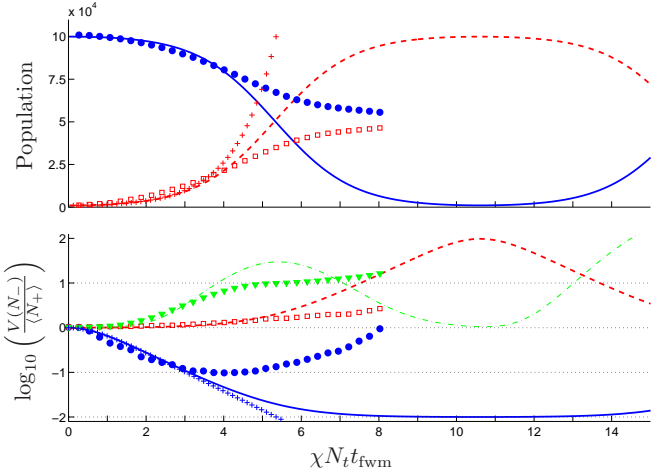}
\caption{\label{fig:populations} (Color online) (a) Populations of each mode as a function of $t_{fwm}$. $\langle \hat{N}_{aL}\rangle$ was calculated using the four-mode TW model (blue solid line), and the 1D multimode TW model (blue solid circles). $\langle \hat{N}_{bL} \rangle$ was calculated using a simple model with no depletion (red ``$+$'' symbols), the four-mode TW model (red dashed line), and the 1D multimode TW model (red squares). $\langle \hat{N}_{aR}\rangle$, and $\langle \hat{N}_{bR}\rangle$ where omitted, as they are almost identical to $\langle \hat{N}_{bL}\rangle$, and $\langle \hat{N}_{aL}\rangle$ respectively. (b) Normalised number difference variance, $v_{i,j}$ for $\{i, j\} = \{aR, bL\}$ (blue ``$+$'' symbols: Simple mode with no depletion; blue solid line: Four-mode TW model; blue solid circles: 1D multimode TW model), $\{aL, bR\}$, (red dashed line: Four-mode TW model; red squares: 1D multimode TW model), and $\{aL, bL\}$ (green dot-dashed line: Four-mode TW model; green triangles: 1D multimode TW model). $\{aR, bR\}$, $\{aL, aR\}$, and $\{bL, bR\}$ are omitted, as they are almost identical to $\{aL, bL\}$. The multimode calculation agrees well with the 4-mode calculation for small values of $t_{fwm}$, but doesn't reach the same degree of squeezing. }
\end{figure}

\fig{fig:populations} shows $\langle \hat{N}_{aR}(t) \rangle$ and $v_{aR, bL}$ as a function of time. $v_{aR, bL}$ decreases from $1$ exponentially with $r$. However, the population grows exponentially, suggesting that our model is invalid for long times, and a model which takes into account depletion is required. 

A more complicated analysis of our system which includes effects such as quantum depletion and Kerr-dephasing due to uncertainties in the particle number can be achieved by using a stochastic phase-space method. Specifically, we use the truncated Wigner (TW) approach \cite{norrie}, which we will now describe. The master equation for the system is found from \eq{4mode_ham}, and then converted into a Fokker-Plank equation (FPE) by using the Wigner representation. This equation can then be converted into a set of stochastic differential equations, which we solve numerically. By averaging over many trajectories with different noises in the initial conditions, expectation values of the quantum field operators can be calculated. When converting our FPE to stochastic differential equations, we ignore terms with third-order derivatives in the FPE, as these terms don't have a simple mapping to the stochastic differential equations, and can be assumed to be negligible when the field has high occupation numbers \cite{norrie}. The stochastic differential equations are
\begin{eqnarray}
i\dot{\alpha}_0 &=& \left(\omega_0 + \chi N_t\right) \alpha_0 + \chi\beta_k^*\beta_0\alpha_k \\
i\dot{\alpha}_{k_0} &=& \left(\omega_{k_0}  +\chi N_t\right)\alpha_k+ \chi\beta_0^*\beta_k\alpha_0 \\
i\dot{\beta}_0 &=& \left(\omega_0 + \chi N_t\right)\beta_0+ \chi\alpha_k^*\alpha_0\beta_k \\
i\dot{\beta}_{k_0} &=& \left(\omega_{k_0} +  \chi N_t\right)\beta_k + \chi\alpha_0^*\alpha_k\beta_0
\end{eqnarray}
where $N_t = \alpha_0^*\alpha_0 + \alpha_{k_0}^*\alpha_{k_0} + \beta_0^*\beta_0 + \beta_{k_0}^*\beta_{k_0} - 2$ represents the total number of particles. The subtraction of $2$ is required due to the correspondence between the complex variables in the stochastic differential equations, and symmetrically ordered operators. These complex variables correspond to our original operators: $\ahat_j \rightarrow \alpha_j$, $\ahatd_j \rightarrow \alpha_j^*$, $\bhat_j \rightarrow \beta_j$, $\bhatd_j \rightarrow \beta_j^*$. The noise on the initial conditions for each trajectory of the evolution these equations was chosen such that they correspond to the quantum statistics of the initial state of interest. Expectation values of symmetrically ordered combinations of field operators correspond to the ensemble average of these complex variables over a large number of trajectories. For example: $\overline{|\alpha_j|^2} = \frac{1}{2}\langle \ahatd_j\ahat_j + \ahat_j\ahatd_j\rangle$, where the overline represents the mean of a large number of trajectories. 

We solve our equations of motion choosing our initial conditions as Glauber coherent states for each mode, with $\langle \hat{N}_{aL}(0)\rangle=\langle \hat{N}_{bR}(0)\rangle  = 10^5$, and $\langle \hat{N}_{aR}(0)\rangle=\langle \hat{N}_{bL}(0)\rangle  = 10^3$. We let the system evolve for an amount of time $t_{fwm}$. In practice, $t_{fwm}$ is determined by the separation time for the two momentum components, and can be adjusted by manipulation of the longitudinal trapping frequency. Alternatively, $\chi$ can be adjusted by manipulation of the scattering length via a Feshbach resonance, or adjusting the transverse trapping frequency.  Figure (\ref{fig:populations}) shows the populations of each mode as a function of $t_{fwm}$. The four-wave mixing creates correlations in particle number between modes $|a,k_0\rangle$ and $|b,0\rangle$, which we quantify by examining the normalized number difference variance $v_{aR, bL}$. As $t_{\mathrm{fwm}}$ increases, $v_{aR, bL}$ initially decreases exponentially until the effects of quantum depletion limit the amount of squeezing. A minimum value of $v_{aR, bL} \approx 10^{-2}$ is reached at $N_t\chi t_{\mathrm{fwm}} \approx 10$, before the correlations begin to decrease again. All other binary combinations of modes display anti-correlations as a result of the four-wave mixing. The degree by which these quantum correlations can enhance an interferometric measurement are show in Fig (\ref{sigvar1}).

\section{Multi-mode effects}\label{sec4}
To investigate the validity of the approximations we made in the previous section, we introduce a one-dimensional model to investigate the effect of multi-mode dynamics on the four-wave mixing process. Assuming the dynamics in the $y$ and $z$ dimensions is trivial, we reduce \eq{fullham} to one dimension by integrating over the $y$ and $z$ dimensions. We then derive a set of stochastic partial differential equations for the complex functions $\psi_i(x)$ corresponding to the quantum operator $\psihat_i(x)$ via the same method as described in the previous section \cite{ferrisFWM}.
\begin{eqnarray}
i\hbar\frac{d}{dt} \psi_a &=& \left(\frac{-\hbar^2}{2m}\frac{\partial^2}{\partial x^2} + U_{1d} n_t(x)\right)\psi_a(x) \label{mm_eom_1} \\
i\hbar\frac{d}{dt} \psi_b &=& \left(\frac{-\hbar^2}{2m}\frac{\partial^2}{\partial x^2} + U_{1d} n_t(x)\right)\psi_b(x) \label{mm_eom_2}
\end{eqnarray}
where $n_t(x) = \left(|\psi_a(x)|^2 -\frac{1}{2\Delta x}\right) + \left(|\psi_b(x)|^2 -\frac{1}{2\Delta x}\right)$ is the total density, where $\Delta x$ is the grid spacing. The subtraction of $\frac{1}{2\Delta x}$ is to compensate for the mean field of the vacuum, which is non-zero in the truncated Wigner approach. $U_{1d} = \frac{U_0}{\pi r_0^2}$ is the one-dimensional interaction strength, and $r_0$ is the characteristic transverse width of the condensate. The initial condition is chosen to be analogous to the initial condition in the four-mode model: 
\begin{eqnarray}
\psi_a(x, 0) &=& \sqrt{10^5}\Psi_0(x) + \sqrt{10^3}\Psi_0(x) e^{ik_0 x} + \frac{\eta_1(x)}{\sqrt{\Delta x}} \nonumber \\
\psi_2(x, 0) &=& \sqrt{10^3}\Psi_0(x) + \sqrt{10^5}\Psi_0(x) e^{ik_0 x} + \frac{\eta_2(x)}{\sqrt{\Delta x}} \nonumber
\end{eqnarray}
where $\eta_j(x)$ are complex Gaussian noise functions with standard deviation in the real and imaginary components of one half and $\langle \eta_n^*(x_i)\eta_m(x_j)\rangle = \frac{1}{2}\delta_{i,j}\delta_{n,m}$. In the TW method, the expectation value of the density of state $|j\rangle$ atoms is $\langle \psihatd_j(x)\psihat_j(x)\rangle = \overline{|\psi_j(x)|^2} -\frac{1}{2\Delta x}$.  At $t=0$, confinement in the $x$-direction is switched off, whilst keeping the transverse confining potential, and the the two wavepackets are allowed to spatially separate. After an amount of time $t_{fwm}$, the $s$-wave scattering length is switched to zero using a feshbach resonance, to control the amount of four-wave mixing that will occur. After $70$ ms, the wavepackets have completely separated, and we measure the number of atoms on either side of a central position $x_0$. Using this measurement, we can define the quantities analogous to the populations of the four modes in the previous section as
\begin{eqnarray}
\hat{N}_{aL} &\equiv& \int_{-\infty}^{x_0}\psihatd_a(x)\psihat_a(x)\, dx \\
\hat{N}_{bL} &\equiv& \int_{-\infty}^{x_0}\psihatd_b(x)\psihat_b(x)\, dx \\
\hat{N}_{aR} &\equiv& \int_{x_0}^{\infty}\psihatd_a(x)\psihat_a(x)\, dx \\
\hat{N}_{aL} &\equiv& \int_{x_0}^{\infty}\psihatd_b(x)\psihat_b(x)\, dx 
\end{eqnarray}

\begin{figure}
\includegraphics[width=0.9\columnwidth]{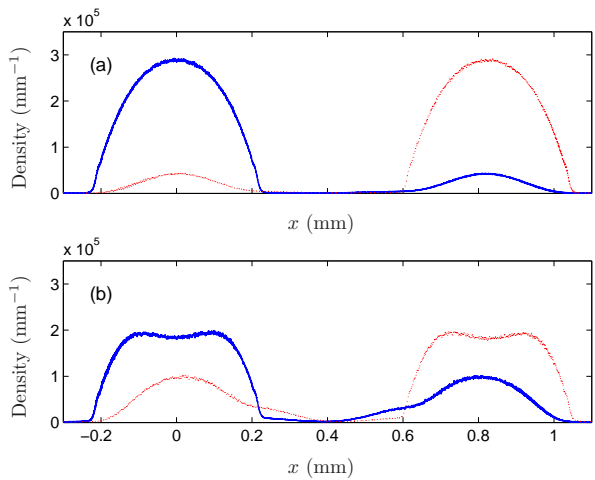}
\caption{\label{fig:densities} $\langle \psihatd_a(x)\psihat_a(x)\rangle$ (blue solid line) and $\langle \psihatd_b(x)\psihat_b(x)\rangle$ (red dotted line) after $70$ ms of separation time for (a): $t_{fwm} = 0.1$ms, and (b): $t_{fwm}=0.18$ms. The expectation values were calculated from $1200$ trajectories. The trapping frequencies of the harmonic potential at $t=0$ were $\{\omega_x, \omega_y, \omega_z\} = 2\pi\{5,1000,1000\}$Hz. The interaction strength was scaled by the cross-sectional area of the condensate: $U_{1d} = \frac{U}{\pi r_0^2}$, with $r_0 =0.55$ $\mu$m. $x_0$ was chosen as the point $x=0.4$ mm, which is roughly halfway between the two wavepackets.}
\end{figure}

\fig{fig:populations} (a) shows the populations in each of these modes as a function of $t_{fwm}$. For early times $N_t\chi t_{fwm} \lesssim 4$ ($t_{fwm} \lesssim 150$ $\mu$s), the multimode model mimics the simple four-mode model. This is also the time when the variances (\fig{fig:populations} (b)) begin to deviate from the simple four-mode model and $v_{aL, bR}$ reaches its minimum value $v_{aL, bR} \approx 0.1$, almost a factor of ten away from the minimum predicted from the simple four-mode model. The discrepancy is due to complications arising from multimode effects. \fig{fig:densities} shows the expectation value of the density of the atomic clouds after separation for $t_{fwm} =0.1$ ms ($N_t\chi t_{fwm} \approx 2.68$)  and $t_{fwm} = 0.18$ ms ($N_t \chi t_{fwm} \approx 4.82$). In (a), the density profile of all wavepackets closely resembles that of the initial state, suggesting that \eq{smode_approx} is a good approximation in this case. However, (b) shows significant deviation from these spatial modes. One cause of this is that the density is higher in the center of each cloud, so the four-wave mixing occurs at a greater rate there, increasing the depletion. It is interesting to note that discrepancies begin to arise in all three models at roughly the same point, as the quantum depletion which causes the un-depleted pump model to break down also causes approximation in \eq{smode_approx} to break down, as the depletion alters the shape of the spatial mode. The 2nd effect is the finite momentum spread of the wavepacket allows for spontaneous scattering to unpopulated momentum modes, which slowly grow. This is the cause of the `tails' leading out from the condensate in \fig{fig:densities} (b). Without the $1\%$ `seed' in the lowly occupied modes, we observe a greater degree of this spontaneous scattering. However, increasing the initial population in the seed modes degrades the degree of quantum correlations. Picking the optimum level of initial population in these modes is a trade-off between suppressing spontaneous scattering into unpopulated modes, and degrading the degree of quantum correlations produced. We did not do a comprehensive search to find the optimum value of this initial population, but found that the model behaved to our satisfaction with a $1\%$ seed. This effect of spontaneous scattering is not present in a semiclassical model, as spontaneous scattering is forbidden. 

\section{Interferometry below the standard quantum limit}\label{sec5}
\begin{figure}
\includegraphics[width=0.75\columnwidth]{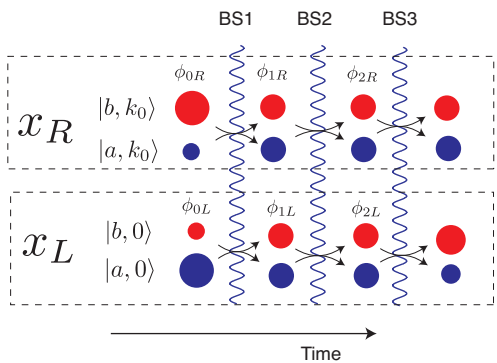}
\caption{\label{fig:pulse_seq} Pulse sequence. After the two momentum modes have spatially separated, the interferometry stage is implemented by a sequence of three $\frac{\pi}{2}$ pulses. An appropriate phase shift $\phi_{0L}$ is applied to $|b,0\rangle$ and $\phi_{0R}$ to $|b,k_0\rangle$ before the 1st coupling pulse and $\phi_{1L}$ and $\phi_{1R}$ before the 2nd pulse to ensure that the populations in each mode remain approximately equal after each pulse. These phase shifts may be different, and can be implemented using the AC Stark shift from a localised optical beam. The phase shifts $\phi_{2L}$ and $\phi_{2R}$ between the 2nd and 3rd pulses is the quantity that the interferometer measures. The population in each mode is measured after the 3rd beam splitter. }
\end{figure}

The obvious method for using the state generated from the four-wave mixing process to perform phase measurements with sensitivity greater than the standard quantum limit would be to use $|a,k_0\rangle$ and $|b,0\rangle$ as the inputs to an atom interferometer, as these are the modes which display two-mode squeezing. However, there are two drawbacks to this approach. The first is that each of these modes are only lowly occupied, so although they would display sub-shot noise sensitivity when used in the appropriate way for an interferometric measurement, the absolute sensitivity of the measurement would be low, as the total number of particles is small. The second draw-back is that these modes aren't spatially overlapping, so a complicated sequence of momentum-shaping pulses would be required in order to construct an atom interferometer out of these two wavepackets. By combining these two lowly occupied modes with $|a,0\rangle$ and $|b,k_0\rangle$, which have a much higher occupation, we may be able to maintain the advantage of the quantum squeezing, while increase the total number of particles, such that the absolute sensitivity is increased. Additionally, $|a,0\rangle$ and $|b,0\rangle$ form a spatially overlapping pair, as do $|a,k_0\rangle$ and $|b,k_0\rangle$. We can combine these four modes to make an interferometric measurement in the following way. The interferometry stage is implemented by a sequence of $\frac{\pi}{2}$ microwave pulses resonant with the $|a\rangle\rightarrow|b\rangle$ transition, which provides local coupling for the spatially overlapping wavepackets. Our scheme is described in \fig{fig:pulse_seq}.  In total, three coupling pulses are used, with a phase shift on $|b,0\rangle$ and $|b, k_0\rangle$ before each coupling pulse. Phase shifts on the left and right wavepacket need to be controlled independently. This could be implemented by the AC Stark shift from independent focussed optical beams. The first coupling pulse is used to prepare the state such that it is useful for atom interferometry, with the phase shift before the first pulse chosen such that the populations in $|a,0\rangle$ and $|b,0\rangle$, and $|a, k_0\rangle$ and $|b, k_0\rangle$ are approximately equal after the first beam splitter. At this stage, the quantity $\hat{S} \equiv (\hat{N}_{aL}-\hat{N}_{bL}) - (\hat{N}_{bR}-\hat{N}_{aR})$ is squeezed. That is, the variance in this quantity is much less than for uncorrelated sources. The phase shift before the second coupling pulse is again chosen such that the populations remain equal. At this point, $\hat{S}$ is anti-squeezed, and the relative phase will be squeezed. The physical process of interest causes a phase shift $\phi_{2L}$ between $|a,0\rangle$ and $|b,0\rangle$, and $\phi_{2R}$ between $|a, k_0\rangle$ and $|b, k_0\rangle$ to be accumulated during the free evolution between the 2nd and 3rd coupling pulses. After the 3rd coupling pulse, the the number of atoms in each mode is measured, and $S$ is calculated. 

For any interferometric device, the smallest detectable phase shift is
\begin{equation}
\Delta \phi = \frac{\sqrt{V(S)}}{\abs{\frac{d \langle \hat{S}\rangle}{d \phi}}} \, , \label{dowlingeq}
\end{equation}
where $S$ is the measured signal. The ultimate limit for sensitivity is the Heisenberg limit $\Delta \phi = \frac{1}{N_t}$, where $N_t$ is the total number of detected atoms. However, when using uncorrelated atoms as the input to the interferometer, the maximum sensitivity is given by the standard quantum limit, $\Delta \phi = \frac{1}{\sqrt{N_t}}$ \cite{dowling}. 

\begin{figure}
\includegraphics[width=0.9\columnwidth]{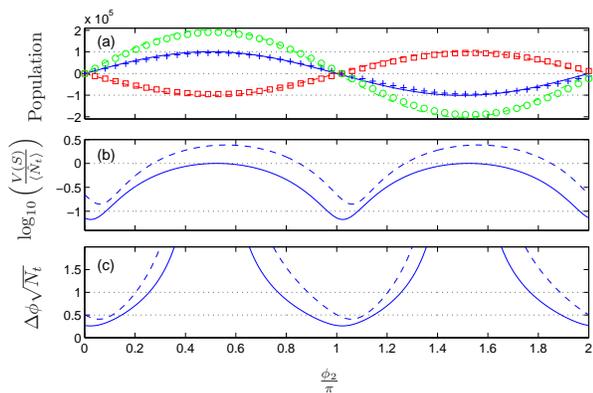}
\caption{\label{sigvar1} (a): $\langle \hat{N}_{aL}-\hat{N}_{bL}\rangle$ (blue solid line: four-mode TW model; blue ``$+$" symbols: 1D multimode TW model), $\langle \hat{N}_{bR}-\hat{N}_{aR}\rangle$ (red dashed line: four-mode TW model; red: 1D multimode TW model), and $\langle \hat{S}\rangle$ (green dot-dashed line: four-mode TW model; green circles: 1D multimode TW model) as a function of $\phi_2$. The four-mode model has slightly more fringe visibility than the 1D multimode model. The horizontal black dotted lines are a visual guide only. (b): $\frac{V(S)}{\langle \hat{N}_t\rangle}$ from the four-mode model (solid blue line), and the multi-mode model (dashed blue line) as a function of $\phi_2$. The  horizontal black dotted lines indicated the level expected for uncorrelated atoms ($\log_{10}\left(  \frac{V(S)}{\langle \hat{N}_t\rangle}\right)=0$), and squeezing below the uncorrelated level by a factor of $10$ ($\log_{10}\left(  \frac{V(S)}{\langle \hat{N}_t\rangle}\right)=-1$). (c) The phase sensitivity $\Delta \phi$ from the four-mode model (blue solid line) and multi-mode model (blue dashed line). The minimum of $\Delta \phi \sqrt{N_t}$ for the multimode model is $\Delta\phi \approx 0.41\sqrt{N_t}$, more than twice as sensitive than for uncorrelated atoms. }
\end{figure}

After the wavepackets have spatially separated, we essentially have two interferometers; one centred at $x_L$, and the other centred at $x_R$. When treating our two interferometers separately, the sensitivity of each device is well above the standard quantum limit. However, when taking our signal to be the difference in the two signals: $\hat{S} = (\hat{N}_{aL} - \hat{N}_{bL}) - (\hat{N}_{bR} - \hat{N}_{aR})$, it may be possible to observe an enhancement in sensitivity, as we have correlations between the left an right systems. We will consider the case where the same applied phase $\phi_2$ is applied to the left and right interferometers, with a constant offset of $\pi$ added to the right interferometer, ie $\phi_{2L} = \phi_2$, $\phi_{2R} = \pi +\phi_2$. In this configuration, the system is set up to measure some quantity that effects each interferometer equally, such as a homogenious magnetic field. By removing the phase shift $\pi$ from the right interferometer, the system is now configured to measure the \emph{difference} in the external phase shifts, and is suited to measure some quantity which is different for each interferometer, such as detecting the difference in magnetic field at two points. \fig{sigvar1} shows $\langle \hat{S}\rangle$, $\frac{V(S)}{\langle \hat{N}_t\rangle}$, and $\Delta \phi$ as a function of $\phi_2$ after the 3rd coupling pulse for $t_{fwm} = 0.12$ms. The minimum of $\Delta \phi \sqrt{N_t}$ for the multimode model is $\Delta\phi\sqrt{N_t} \approx 0.41$, more than twice as sensitive than for uncorrelated atoms. The fringe visibility of the final interferometric measurement is $\sim 0.999$ for the four mode model, and $\sim 0.935$ for the multimode model.   

\section{Comparison with one-axis twisting}\label{sec6}
An alternate scheme for performing atom interferometry with sub-shot noise sensitivity is the previously discussed one-axis twisting scheme \cite{oberthaler2010, treutlein2010, ueda, hainekerr}. This scheme involves splitting the condensate, and allowing the nonlinear interactions to `shear' the relative phase of the two modes until the desired level of spin squeezing is reached. However, at this point, the squeezing is not directly usable, and the quadrature squeezing needs to be rotated by an angle $\theta$ such that there are reduced fluctuations in the relative phase. In order to achieve this, an additional phase shift and a coupling pulse is applied, in an analogous method as discussed in section \ref{sec5} of this article. Significant anti-squeezing instead of squeezing will be observed if the incorrect phase shift is applied. This phase shift depends critically on the total number of particles, as the nonlinear effect which causes the phase shearing also causes a relative phase shift between the two modes which depends on the number of particles. As the number of particles becomes large, a small variation in the total number of particles is sufficient to significantly alter the relative phase of the two modes, and thus rotate the squeezing quadrature such that anti-squeezing is observed. 

To compare the performance of the one-axis twisting scheme to our four-wave mixing scheme, we analysed the one-axis twisting scheme via a two-mode model described in \cite{hainekerr}. To demonstrate sub-shotnoise sensitivity, one would have to find the optimum value of the phase shift, and then perform many shots of the experiment to build up quantum statistics in the signal. 
We set the total number of particles as $2\times10^5$, and optimising the nonlinear interaction time, coupling pulses, and phase shifts, such that the phase sensitivity of the device was $\Delta \phi\sqrt{N_t} \approx 0.4$. We then varied the total number by a small amount, whilst keeping all other parameters (phase shifts and coupling pulses) fixed. We found that by altering the total number of particles by about $1\%$ was enough to decrease the sensitivity to the limit set by shot-noise ($\Delta \phi = \frac{1}{\sqrt{N_t}}$). It is important to note that this degradation of the signal is not due simply to the fluctuations in the number of atoms directly adding noise to the signal, as the signal (ie, difference in the number of atoms in each mode) is insensitive to fluctuations in the total number of atoms. A small shift in total number causes a big shift to the value of the phase shift required in order to rotate the squeezing to the correct quadrature. However, when the total number of atoms was small ($\sim 1000$), as in the recent experiments by \cite{oberthaler2010, treutlein2010}, a small fluctuation in total number causes a relatively small phase shift, and a perturbation in the number of particles of $\sim 35\%$ was required in order to decrease the sensitivity to the shot-noise limit.

As the Hamiltonian for the fourwave mixing process doesn't rely on asymmetric scattering lengths between the two modes, the process that produces the correlation does not also cause a number dependent phase shift.  We found that altering the initial number by $50\%$ in a simple four-mode calculation, while keeping all other parameters the same as was used in \fig{sigvar1}  ($\chi$, $t_{fwm}$, $\phi_{0R}$, $\phi_{1R}$, $\phi_{2R}$, $\phi_{0L}$, $\phi_{1L}$, and $\phi_{2L}$), the shot-noise limit was still surpassed.  This indicates that four-wave mixing is sufficiently robust against shot-to-shot number fluctuations that it can be used with a large number of particles.

\section{Conclusion}
We have demonstrated that quantum correlations generated from four-wave mixing can be utilised to perform atom interferometry with sensitivity greater than the standard quantum limit. Over longer timescales, multi-mode effects cause complicated evolution in the spatial mode of the wavepackets, which reduce the strength of the useful correlations and ultimately limit the amount of squeezing once can produce. By combining lowly occupied, yet correlated modes with the two remaining highly occupied modes, it is possible to surpass the shot-noise limit, while still using a large number of atoms, increasing the absolute sensitivity of the device.

\section{Acknowledgments}
We would like to acknowledge useful discussions with Matthew Davis, Tod Wright, and Jacopo Sabatini. This work was supported by the Australian Research Council Centre of Excellence for Quantum-Atom Optics, and by Australian Research Council discovery project DP0986893.


\begin{thebibliography}{99}
\bibitem{pritchard_review} Cronin, Alexander D. and Schmiedmayer, J\"org  and Pritchard, David E., Rev. Mod. Phys. {\bf 81},  1051 (2009).
\bibitem{chu99} A. Peters, K. Y. Chung, and S. Chu, Nature {\bf 400}, 849 (1999).
\bibitem{chu2001} A. Peters, K. Y. Chung, and S. Chu, Meteroliga, {\bf 38}, 25, (2001).
\bibitem{kasevich02} J. M. McGuirk, G. T. Foster, J. B. Fixler, M. J. Snadden, and M. A. Kasevich, Phys. Rev. A {\bf 65}, 033608 (2002).
\bibitem{kasevich97} T. L. Gustavson, P. Bouyer, and M. A. Kasevich, Phys. Rev. Lett. {\bf 78}, 2046 (1997).
\bibitem{stamperkurn07} M. Vengalattore, J. M. Higbie, S. R. Leslie, J. Guzman, L. E. Sadler, and D. M. Stamper-Kurn, Phys. Rev. Lett. {\bf 98}, 200801 (2007).
\bibitem{kasevichG} J. B. Fixler, G. T. Foster, J. M. McGuirk, and M. A. Kasevich, Science {\bf 315}, 5808 (2007).
\bibitem{biraben_alpha} R. Bouchendira, P. Clade, S. Guellati-Khelifa, F. Nez, F. Biraben, Phys. Rev. Lett. {\bf 106}, 080801 (2011).
\bibitem{close2011} J. E. Debs, P. A. Altin, T. H. Barter, D. D\"{o}ring, G. R. Dennis, G. McDonald, R. P. Anderson, N. P. Robins, and J. D. Close, arXiv:1011.5804v3 (2011).
\bibitem{dowling} J. P. Dowling, Phys. Rev. A {\bf 57}, 4736 (1998).
\bibitem{oberthaler2010} C. Gross, T. Zibold, E. Nicklas, J. Esteve, and M. K. Oberthaler, Nature {\bf 464}, 1165 (2010).
\bibitem{treutlein2010} M. F. Riedel, P. Bohl, Y. Li, T. W. Hansch, A. Sinatra, and P. Treutlein Nature {\bf 464}, 1170 (2010).
\bibitem{ueda} M. Kitagawa and M. Ueda, Phys. Rev. A {\bf 47}, 5138 (1993).
\bibitem{hainekerr} S. A. Haine and M. T. Johnsson, Phys. Rev. A {\bf 80}, 023611 (2009). 
\bibitem{phillipsFWM} L. Deng, E. W. Hagley, J. Wen, M. Trippenbach, Y. Band, P. S. Julienne, J. Simsarian, K. Helmerson, S. L. Rolston, and W. D. Phillips, Nature, {\bf 398}, 218 (1999).
\bibitem{ketterleFWM1} J. M. Vogels, K. Xu, and W. Ketterle, Phys. Rev. Lett. {\bf 89}, 020401 (2002).
\bibitem{ketterleFWM2} J. M. Vogels, J. K. Chin, and W. Ketterle, Phys. Rev. Lett. {\bf 90}, 030403 (2003).
\bibitem{ferrisFWM} A. J. Ferris, M. K. Olsen, and M. J. Davis, Phys. Rev. A {\bf 79}, 043634 (2009).
\bibitem{trippenbach_collision} P. Zin, J. Chwedenczuk, A. Veitia, K. Rzkazewski, and M. Trippenbach, Phys. Rev. Lett. {\bf 94} 200401 (2005).
\bibitem{drummond_collision} P. Deuar and P. D. Drummond, Phys. Rev. Lett. {\bf 98} 120402 (2007).
\bibitem{ogren2008} M. \"{O}gren and K. V. Kheruntsyan,  Phys. Rev. A {\bf 79}, 021606(R) (2009).
\bibitem{dall2009} R. G. Dall, L. J. Byron, A. G. Truscott, G. R. Dennis, M. T. Johnsson, and J. J. Hope, Phys. Rev. A {\bf 79}, 011601(R) (2009).
\bibitem{dennis2010} G. R. Dennis and M. T. Johnsson, Phys. Rev. A {\bf 82} 033615 (2010).
\bibitem{aspectFWM} A. Perrin, H. Chang, V. Krachmalnicoff, M. Schellekens, D. Boiron, A. Aspect, and C. I. Westbrook, Phys. Rev. Lett. {\bf 99} 150405 (2007). 
\bibitem{aspectFWM2} J.C Jaskula, M Bonneau, G.B Partridge, V Krachmalnicoff, P Deuar, K.V Kheruntsyan, A Aspect, D Boiron, and C.I Westbrook, Phys. Rev. Lett. {\bf 105} 190402 (2010). 
\bibitem{pertot} D. Pertot, B. Gadway, and D. Schneble, Phys. Rev. Lett. {\bf 104}, 200402 (2010).
\bibitem{Walls} Walls D. F. and Milburn G. J., Quantum Optics (Berlin: Springer) (1994).
\bibitem{norrie} A. A. Norrie, R. J. Ballagh, and C. W. Gardiner, Phys. Rev. A {\bf 73}, 043617 (2006). 
\end{thebibliography}
\end{document}